\documentclass{mn2e}
\input epsf.sty

\def\Rg{R_{\rm g}}

\def\lh{l_{\rm h}}

\def\lsoft{l_{\rm s}}
\def\OmegaK{\Omega_{\rm K}}

\def\vphi{v_{\phi}}
\def\vz{v_{\rm z}}
\def\tauf{\tau_{\rm fl}}

\def\tbin{t_{\rm bin}}
\def\tmin{\tau_{\rm min}}
\def\tmax{\tau_{\rm max}}
\def\trise{\tau_{\rm r}}
\def\tdecay{\tau_{\rm d}}

\def\rms{R_{\rm ms}}
\def\rin{R_{\rm in}}

\def\phis{\phi_{\rm s}}

\def\sigmaT{\sigma_{\rm T}}
\def\me{m_{\rm e}}

\def\MSun{{\rm M}_{\odot}}

\def\Ka{K$\alpha$\ }
\def\Pt{P_{\tau}}

\title[Relativistic effects in X-ray variability]
{On the influence of relativistic effects on X-ray variability of accreting 
 black holes}

\author[P. T. \.{Z}ycki and A. Nied\'{z}wiecki]{Piotr T. \.{Z}ycki$^1$\thanks{e-mail: ptz@camk.edu.pl} and Andrzej Nied\'{z}wiecki$^2$ \\
    $^1$Nicolaus Copernicus Astronomical Center, Bartycka 18, 00-716 Warsaw, Poland \\
    $^2${\L}\'{o}d\'{z} University, Department of Physics, Pomorska 149/153, 90-236 {\L}\'{o}d\'{z}, Poland}

\date{7 February 2005}

\begin{document}
\label{firstpage}

\maketitle

\begin{abstract}

X-rays produced by compact flares co-rotating with a Keplerian accretion
disc are modulated in time by Doppler effects.
We improve on previous calculations of these effects by considering
recent models of intrinsic X-ray variability, and compute the expected
strength of the relativistic signal in current data of Seyfert
galaxies and black hole binaries. Such signals could clearly be seen
in, for example, recent {\it XMM-Newton\/} data from MCG--6-30-15, 
if indeed the X-rays
were produced by co-rotating flares concentrated toward the inner disc edge 
around an extreme Kerr black hole.
Lack of the signal in the data collected so far gives support to models,
where the X--ray sources in active galaxies do not follow Keplerian orbits 
close to the black hole.

\end{abstract}


\section{Introduction}

Matter accreting onto compact objects (black holes, neutron stars) can be
accelerated to
relativistic velocities. The best observational evidences of
such velocities come from  distortions of spectral features
due to Doppler energy shifts. One of the best evidences for relativistic
motion in accreting black holes (both in Active Galactic Nuclei, AGN, and
black hole binaries, BHB), are the broad \Ka lines of Fe near 6.5 keV
(review in Reynolds \& Nowak 2003). These can be explained by Doppler
and gravitational energy shifts, from Keplerian motion of the emitting
plasma, deep in the gravitational potential wells of the central objects.

Spectral distortions are not the only effect of relativistic motions of
sources of radiation. The observed flux should be modulated in
time, corresponding to periodic Doppler enhancement and reduction as the source
revolves around the central body. The primary emission will be modulated,
if its source also participates in the rapid Keplerian motion.
We can thus expect periodic time modulation
of the primary X-ray emission from accreting  compact objects in those spectral states,
when the emission is coming from magnetic flares. This applies to, for example,
BHB in soft states. Generally, such time modulation should be observed
from sources showing broad Fe lines, again, under the assumption that
the primary emission originates in magnetic flares rotating with
the disc, rather than, e.g., a base of a jet.
The presence of a magnetically driven active corona is naturally expected from 
evolution of magnetic field inside an accretion disc (e.g.\ Galeev, 
Rosner \& Vaiana 1979), and it is in fact supported by recent magnetohydrodynamical
simulations (Turner 2004). 

The purpose of this paper is to estimate the strength
of the modulation in X-ray power density spectra (PDS) from accreting
black holes. 
First computations of the appearance of a source orbiting a black hole
were performed by Cunningham \& Bardeen (1972) and Bardeen \& Cunningham (1973)
The influence of relativistic effects on time variability of accreting black
holes was estimated
by Abramowicz et al.\ (1991) and subsequent papers, e.g., Bao (1992),
Bao \& Ostgaard (1995).
They assumed  a collection of uncorrelated emitting hot spots, so the variability
came in part from the emergence and disappearance of the hot spots, and
in part from the relativistic effects.
We are more specific in our approach as to the form of the intrinsic 
(i.e.\ before the relativistic effects) variability.

We assume that the basic form of variability PDS is
realized {\em without\/} the relativistic effects (we use the flare
avalanche model of Poutanen \& Fabian 1999; see below). The latter give
additional observable results.
The motivation for formulating the basic model without relativistic effects
is that the observed PDS has the basic form of (at least approximately) 
a doubly 
broken power law for many kinds and spectral states of accreting compact
objects. 
This includes both the low/hard and high/soft states of black hole X-ray 
binaries,  where the accretion flow geometries seems to be quite different, 
yet the power
spectra are similar, at least in their high-frequency parts
(reviews in Done 2002, Zdziarski \& Gierli\'{n}ski 2004). The X-ray power
spectra of Seyfert galaxies have the same form, with break frequencies
roughly scaling with central mass (Czerny et al.\ 2001; Markowitz et al.\ 2003).
Also, the
time scales of broad band X-ray variability in accreting compact systems 
are much longer than the Keplerian time scales of the inner accretion disk,
so other mechanisms of generating the variability must operate.
We note that Miniutti \& Fabian (2004) attribute much of the complex X--ray 
variability
of MCG--6-30-15 specifically to Doppler and light bending effects from close 
to the
central black hole. At the same time, Vaughan, Fabian \& Nandra (2003) 
noticed similarity of
power spectra, time lags and coherence function between MCG--6-30-15 and
Cyg X-1, with a simple scaling with the black hole mass.

However, even though the basic variability model does not seem to require 
relativistic
effects, these effects are quite clear in energy spectra of many sources.
In extreme cases, model fits to the observed profiles of Fe \Ka line
indicate that the emission must come from well within
$6\Rg$, the marginally stable orbit in Schwarzschild metric. 
This is claimed to be the case for Seyfert 1 galaxy MCG--6-30-15 
(Fabian et al.\ 2002), black hole binaries GX 339-4 (Miller et al.\ 2004)
and XTE J1650-500 (Miller et al.\ 2002), where the estimated inner radii are
$\approx 2$--3$\Rg$.
In addition, the radial emissivity of the line seems to be strongly peaked 
toward the inner radius. Thus, the relativistic effects should 
contribute to the variability characteristics, for example, the power spectra.

We adopt the flare avalanche variability model of Poutanen \& Fabian (1999;
hereafter PF99), and assign radial locations to the flares. 
We implement the relativistic effects in extreme Kerr metric ($a=0.998$), 
and compute the modulation in the observed X-ray light curves due to
relativistic effects, as a function of inner disc radius and inclination angle.
We compute the PDS, and demonstrate that for the extreme cases of inner
radius being close to the marginally stable orbit ($\rms=1.23\Rg$),
the signal should be observed in current high quality XMM data.

\section{The Model}
\label{sec:model}

\subsection{Variability model and energy spectra}

The model is very similar to that used by \.{Z}ycki (2002), and it is based
on the variability model of PF99.
The X-ray emission is assumed to be produced in active region filled in with
hot plasma, related to, e.g., magnetic activity in accretion disc corona.
An active region produces a flare of radiation by inverse-Compton
upscattering of soft photons from the accretion disc. The soft photons,
in turn, come mostly from reprocessing of the hard X-rays. The spectrum
of emitted radiation evolves in time during the flare, from softer 
to a rather hard one close to the flare peak. This evolution can be caused
by, e.g., rising the active region above the disc during the flare.
The flares are correlated to produce longer ``avalanches'', which help
explaining the $f^{-1}$ part of the power spectrum (see PF99 for a 
detailed description).
Compared to the original formulation of PF99 we modified the shape of the
heating function, $\lh(t)$ (where $l\equiv (L/R_{\rm flare})(\sigmaT/\me c^3)$) 
is the compactness parameter). It is assumed to be a double exponential
function
with the rise time much longer than the decay time, $\trise \gg \tdecay$,
based on the auto- and cross-correlation analysis by Maccarone, Coppi \& 
Poutanen (2000). The rise time scales of the flares are distributed
between $\tmin$ and $\tmax$ with the distribution $\Pt(\tau) \propto \tau^{-p}$
(see PF99). The cooling of the hot plasma is by soft disc photons from
reprocessed X-rays, $\lsoft(t) \equiv \lh(t) D(t)$, and the feedback 
function, $D(t)$,  is defined as in PF99.

We note that the model of PF99 was originally formulated for hard/low
state of BHB, where the PDS has the characteristic shape of a doubly
broken power law. Power spectra in the soft state have somewhat different
form, since the low frequency break (from $f^{0}$ to $f^{-1}$) is not
observed (Cyg X-1, Churazov, Gilfanov \& Revnivtsev 2001; 
NGC 4051 M$^{\rm c}$Hardy et al.\ 2004), at least down to frequency
$\approx 10^{-5}\times f_{\rm h,br}$, where $f_{\rm h,br}$
is the high break frequency. Also, the shape of the break in the soft 
state PDS is more gradual than in hard state.
However, these differences do not affect our reasoning and
computations, because the considered effects appear on the $f^{-2}$ part
of PDS, i.e.\ past the high-frequency break. This part of PDS is formed
by superposition of Fourier transforms of individual flare profiles,
and as such, depends on properties of individual flares and distribution
of their life times. We also note than in the extreme soft state of Cyg X-1
the rollover frequency seems to be similar to that in the hard state
(fig. 4 in Zdziarski \& Gierli\'{n}ski 2004),
i.e.\ there seems to be {\em no\/} difference in the PDS extent between
the hard and soft states in Cyg X-1.

Flares are assumed to follow Keplerian orbits, starting from a randomly 
selected azimuthal location, so that the primary
emission  is modulated by the relativistic and kinematic effects.
The vertical velocity of a rising plasma region (PF99), $\vz$, is usually 
much smaller than the Keplerian velocity,
\begin{equation}
 {\vz \over \vphi} = {H \over R} {1 \over \OmegaK \tauf},
\end{equation}
(where $H$ is the height above the disc of the point of flare's maximum 
luminosity) since $H/R \ll 1$. Therefore we neglect the vertical motion
while computing the Doppler effects.
For the purpose of computing the relativistic effects we assume that the
emission comes from the equatorial plane.

Crucial model assumption is the radial distribution of flares. A steep 
centrally peaked distribution obviously enhances the importance of relativistic
effects, while a flatter distribution makes those effects less important.
We assume here that the flares are distributed in such a way that the energy
emitted in X-rays follows the prescription for gravitational energy release 
in extreme Kerr metric, $Q(r)$ (see, e.g., Krolik 1999, p.\ 152).
The steep emissivity is also motivated by the results of fitting the observed 
profile of the Fe \Ka line, which  in some cases indicate very steep and centrally
peaked emissivity (Fabian et al.\ 2002; Miller et al.\ 2002, 2004).
The above assumption of emissivity is an important change compared
to our previous model (\.{Z}ycki 2002). There the assumption was made 
that the radial location of a flare is related to its time scale, 
$r \propto \tauf^{2/3}$ derived from $\tauf \propto \OmegaK^{-1}$,
as expected from numerical simulations (e.g.\ Romanova et al.\ 1998).
This however leads to contradictions with 
observations, since, with all the assumptions of the original model of PF99, 
the
longer flares emit more energy than the short ones, yet they are located
farther away from the center (see  \.{Z}ycki 2002 for detailed discussion). 
Therefore, in the present model there is no  relation
between the flare time scale and its location, but we also discuss in 
Sec~\ref{sec:depend} if relaxing this assumption can modify our results. 
Radial position of a spontaneous
flare is generated from probability distribution $P_r(r) \propto r Q(r)$
(corresponding to energy dissipation per ring),
and all the flares stimulated by a given spontaneous flares are located
at the same radius. 

At each moment of time the energy spectrum is computed. 
We approximate
the Comptonization spectrum as a sum of a power law with exponential cutoff
and the Wien spectrum, following Zdziarski (1985 and references therein).
The spectral slope and electron temperature  are computed using formulae of 
Beloborodov (1999a,b).
The reprocessed component is not computed. The Comptonized spectrum, $N_{\rm em}$,
is normalized to $\lh$.

\subsection{Effects of photon propagation in Kerr metric}

We consider a Kerr black hole characterized by its mass, $M$, and
angular momentum, $J$. We use the Boyer-Lindquist coordinate
system $(t,R,\theta,\phi)$. 
The following dimensionless parameters are used below
\[
r = {R \over R_{\rm g}},~~~\hat t = {ct \over R_{\rm g}}, 
~~~a = {J \over c R_{\rm g} M},
~~~\Omega = {d\phi \over {d{\hat t}}},
\]
where  $R_{\rm g} = GM/c^2$ is the gravitational radius. We assume the
highest expected value of the spin parameter, $a=0.998$ (Thorne
1974).

We find the photon number flux observed from a source located at $r_{\rm em}$, 
$N_{\rm obs}(r_{\rm em})$, 
by integrating the photon number flux in the source rest frame, $N_{\rm em}$, 
over all initial 
directions for which photons reach the observer, and correcting the photon
emission rate by the time dilation factor, $d\tau / d t$, where $d\tau$ and 
$dt$ are time intervals measured in the rest frame and at infinity, respectively 
(see, e.g., Bardeen 1973). 

All the photon propagation effects in our model are treated by the transfer function, 
${\cal T}$, defined by
\begin{eqnarray}
\lefteqn{ N_{\rm obs}(E,t,i,r_{\rm em})  = } \nonumber \\ & &
   \int {\cal T}(r_{\rm em},i,g,\Delta \hat t) N_{\rm em}
(E/g,t - \Delta t) 
{d\tau \over {d t}} {\rm d} g,
 \label{equ:tf}
\end{eqnarray}
where $i$ is the inclination angle, $\Delta t$ is time interval between emission
and detection of a given bunch of photons, $\Delta t = \Delta \hat t R_g/c$,
and $g$ is the shift of photon energy, $g = E_{\rm obs}/E_{\rm em}$.
Note that the transfer function is independent of $M$. The only dependence 
on the black hole mass occurs in physical scaling of photon travel time and 
disc area. 
Then, total observed spectrum is given by
\begin{equation}
N_{\rm tot}(E,t,i) = \left(\Rg\over D\right)^2 
 \int\limits_{r_{\rm in}}^{r_{\rm out}} N_{\rm obs}(E,t,i,r) Q(r) 2 \pi r {\rm d} r,
\end{equation}
where function $Q(r)$ gives the radial distribution of the emission,
$r_{\rm in}$ and $r_{\rm out}$ are the inner and outer radius of emitting region
($r_{\rm out}$ is fixed at $100\Rg$) 
and $D$ is the distance to the source.

We construct the transfer functions following a Monte Carlo method described in 
detail in Gierli\'{n}ski, Macio{\l}ek-Nied\'{z}wiecki \& Ebisawa (2001). 
However, Gierli\'{n}ski et al.\ (2001) consider time-averaged emission, while our 
model treats time-dependent spectra. Therefore, we extend the procedure 
to include additional information on the time  of photon propagation, as follows.

A large number
of photons are emitted from a point source on a Keplerian orbit in the equatorial
plane. The angular velocity of the source with respect to the distant
observer
\begin{equation}
\Omega_{\rm K}= {1 \over r^{3/2}+a}.
\end{equation}
We assume that the source is isotropic and then
initial directions of photons are generated from the 
distribution uniform in both $\phi_{\rm em}$ and $\cos \theta_{\rm em}$,  
where $\theta_{\rm em}$ is a polar angle between the 
photon initial direction and the normal to the equatorial plane and $\phi_{\rm em}$ 
is the azimuthal angle, in the equatorial plane, with respect to the $r$-direction. 
Solving equations of photon motion, as described below,  we find the energy shift, 
inclination and arrival time, ($g$, $i$, $\Delta \hat t$).

Each element of 
the transfer function, ${\cal T}(r, i, g,\Delta \hat t)$, is computed by 
summing all photon trajectories  emitted from $r$ 
for all angles ($\phi_{\rm em}$, $\theta_{\rm em}$) 
for which required ($g$, $i$, $\Delta \hat t$) are obtained.

The travel time and the change of azimuthal angle of a photon 
emitted at a distance $r$ from the equatorial plane
are given by (e.g.\ Misner, Thorne \& Wheeler 1973)
\begin{equation}
\Delta \phi_{\rm ph} =  \int\limits_{\pi/2}^i {\lambda - a \sin^2 \theta \over 
  \Theta^{1/2} \sin^2 \theta  } {\rm d} \theta + \int\limits_{r}^D
{a(r^2 + a^2 - \lambda a) \over \Delta \Re^{1/2} } {\rm d}r,
\end{equation}
\begin{equation}
\Delta \hat t_{\rm ph} =  \int\limits_{\pi/2}^i {a(\lambda - a \sin^2 \theta) \over 
  \Theta^{1/2} } {\rm d} \theta  + \int\limits_{r}^D
{(r^2 + a^2)(r^2 + a^2 - \lambda a) \over \Delta \Re^{1/2} }  {\rm d}r,
\end{equation}
where $\Re$ and $\Theta$ are radial and polar effective potentials, respectively,
\begin{eqnarray}
\lefteqn{\Re(r) = (r^2+a^2-\lambda a)^2 - \Delta \left[ \xi^2 + (\lambda - a)^2
\right] ,} \nonumber \\
\lefteqn{\Theta(\theta) =\xi^2 + \cos^2 \theta \left( a^2-\lambda^2/{\sin^2 
\theta} \right),}
\end{eqnarray}
$\xi$ and $\lambda$ are photon  constants of motion, related to the emission angles by 
\begin{eqnarray}
\lambda & = & {\sin\theta_{\rm em} \sin \phi_{\rm em}+ V \over (r^2 \Delta^{1/2}+
2 a r V)/ A+\Omega_{\rm K} \sin\theta_{\rm em} \sin \phi_{\rm em}},
 \nonumber \\
\xi & = & \left( {A \over \Delta } \right)^{1/2} (1-
V^2)^{-1/2}(1-\lambda \Omega_{\rm K}) \cos \theta_{\rm em},
\label{const}
\end{eqnarray}
\begin{eqnarray}
 \Delta & = &r^2-2r+a^2 \\
 A & = & (r^2+a^2)^2 - a^2 \Delta \\
 V & = & (\Omega_K - 2ar/A) {A \over {r^2 \Delta^{1/2}}}, 
\end{eqnarray}
and the angle, $i$, at which the photon is observed far from the source  
is determined by the integral equation of motion
\begin{equation}
\int_{r}^D \Re^{-1/2}{\rm d} r=\int_{\pi/2}^{i} \Theta^{-1/2} {\rm d}\theta.
\label{int}
\end{equation}
The shift of energy is given by
\begin{equation}
g=r \left( {\Delta \over A} \right)^{1/2} {(1-V^2)^{1/2}
\over 1-\Omega_{\rm K} \lambda}.
\label{g}
\end{equation}

\begin{figure*}
 \parbox{\textwidth}{
  \parbox{0.45\textwidth}{
   \epsfxsize = 0.4\textwidth
   \epsfbox[18 150 620 710]{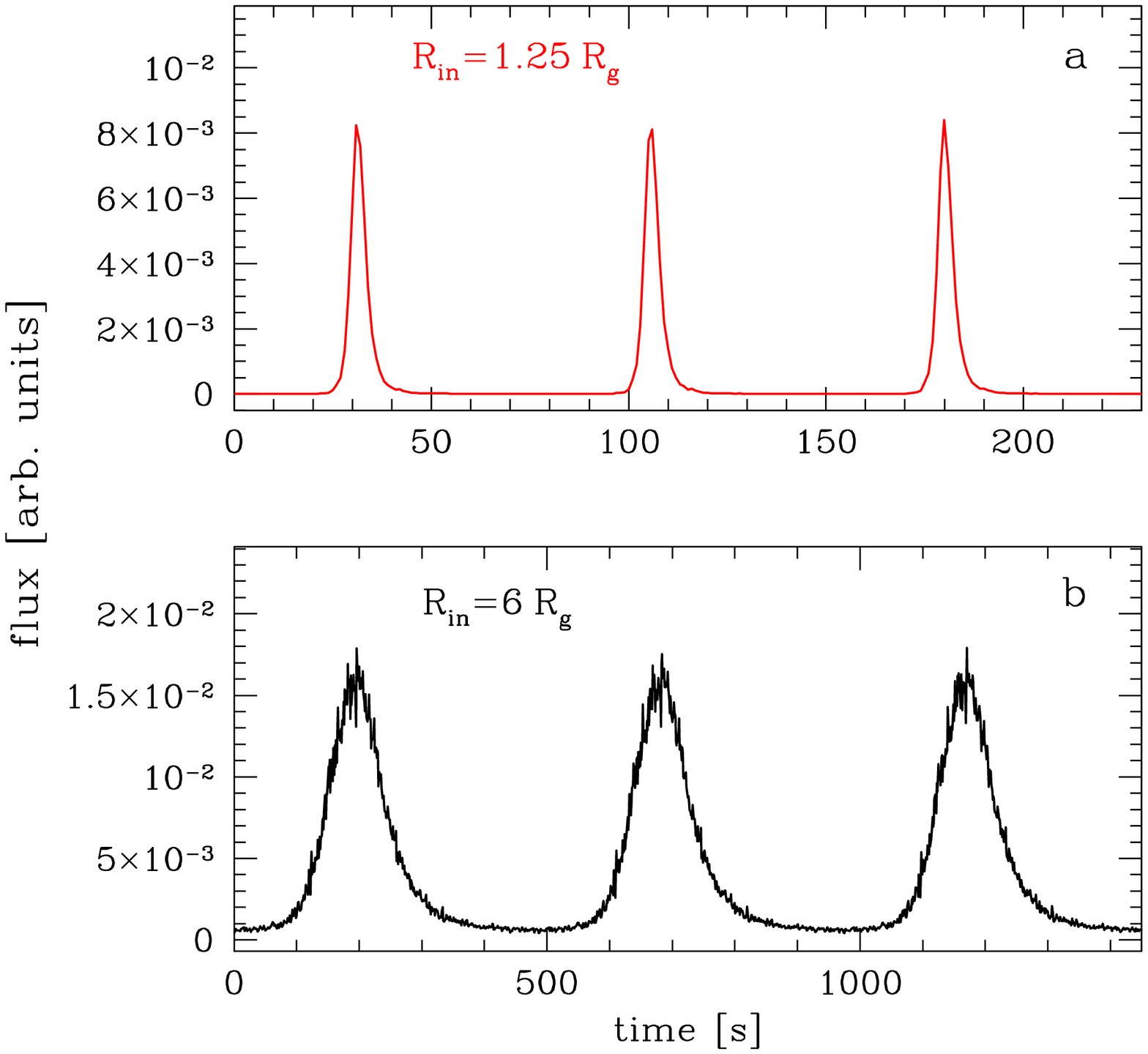}
} \hfil {
 \parbox{0.45\textwidth}{
  \epsfxsize = 0.4\textwidth
  \epsfbox[18 150 620 710]{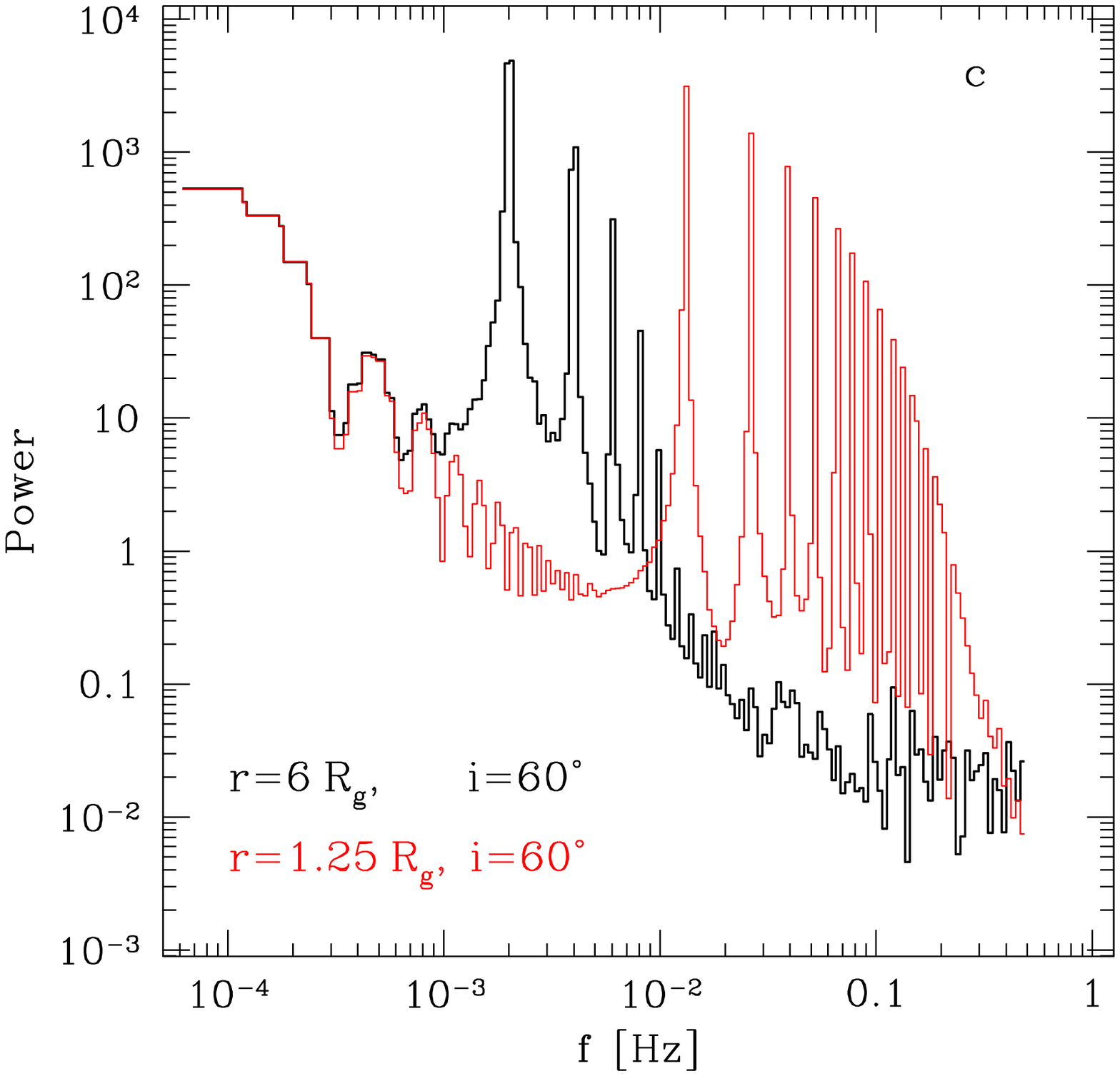}
}
}}
 \caption{
Light curves (a, b) and power spectra (c) from an intrinsically constant 
signal, 
modulated by relativistic effects in Kerr metric with $a=0.998$. 
Emission radius $1.25\Rg$ (a), and $6\,\Rg$ (b), disk inclination $60^{\circ}$.
Black hole mass is $10^6\,\MSun$.
\label{fig:modconst}}
\end{figure*}

A photon with trajectory characterized 
by $\Delta \phi_{\rm ph}$ must be emitted when the source 
is located at $\phis = \phi_{\rm obs} - \Delta \phi_{\rm ph}$
to reach the observer whose azimuthal angle is $\phi_{\rm obs}$.     
Then, taking into account 
$\phi_s (t)= \phi_0 + \Omega_K t$,
where $\phi_0$ is a random initial location,
we find the arrival time 
of the photon $\Delta \hat t = \Delta \hat t_{\rm ph} - \Delta \phi_{\rm ph}/\Omega_K$ for
$\Delta \phi_{\rm ph} < 0$ and 
$\Delta \hat t = \Delta \hat t_{\rm ph} - (2 \pi - \Delta \phi_{\rm ph})/\Omega_K$
for $\Delta \phi_{\rm ph} > 0$.

The time dilation factor is given by
\begin{equation}
 {d\tau \over {d t}} = r \left( {\Delta \over A} \right)^{1/2} (1-V^2)^{1/2}.
\label{equ:dtdt}
\end{equation}
The transfer function for a given radius is normalized to the total number
of emitted photons. This means that in includes the effect of photon capture
by the black hole/accretion disc. This effect suppresses the emission from
close to the black hole reaching a distant observer. 

Our transfer function is not ideally smooth when plotted, for example, 
as a function of $\Delta\phi_{\rm ph}$ for a given radius, which is a consequence 
of finite statistics in the Monte Carlo method.
This is apparent in some cases, at highest frequencies of PDS.

The sequence of spectra created by the above procedure is subject to standard 
analysis in the time and Fourier domains (see e.g.\ van der Klis 1995; 
Nowak et al.\ 1999; Poutanen 2001). The power spectra are computed
from light curves at 3 keV.

All computations are done assuming the central black hole mass $M=10^6\,\MSun$.
The parameters of intrinsic variability model are assumed to follow a simple 
scaling with the black hole mass.
That is, their values are adopted as for Cyg X-1 and then scaled by 
a factor of $10^5$

\section{Results}
 \label{sec:results}

\begin{figure*}
 \parbox{\textwidth}{
   \parbox{0.45\textwidth}{
  \epsfxsize = 0.4\textwidth
  \epsfbox[18 150 620 710]{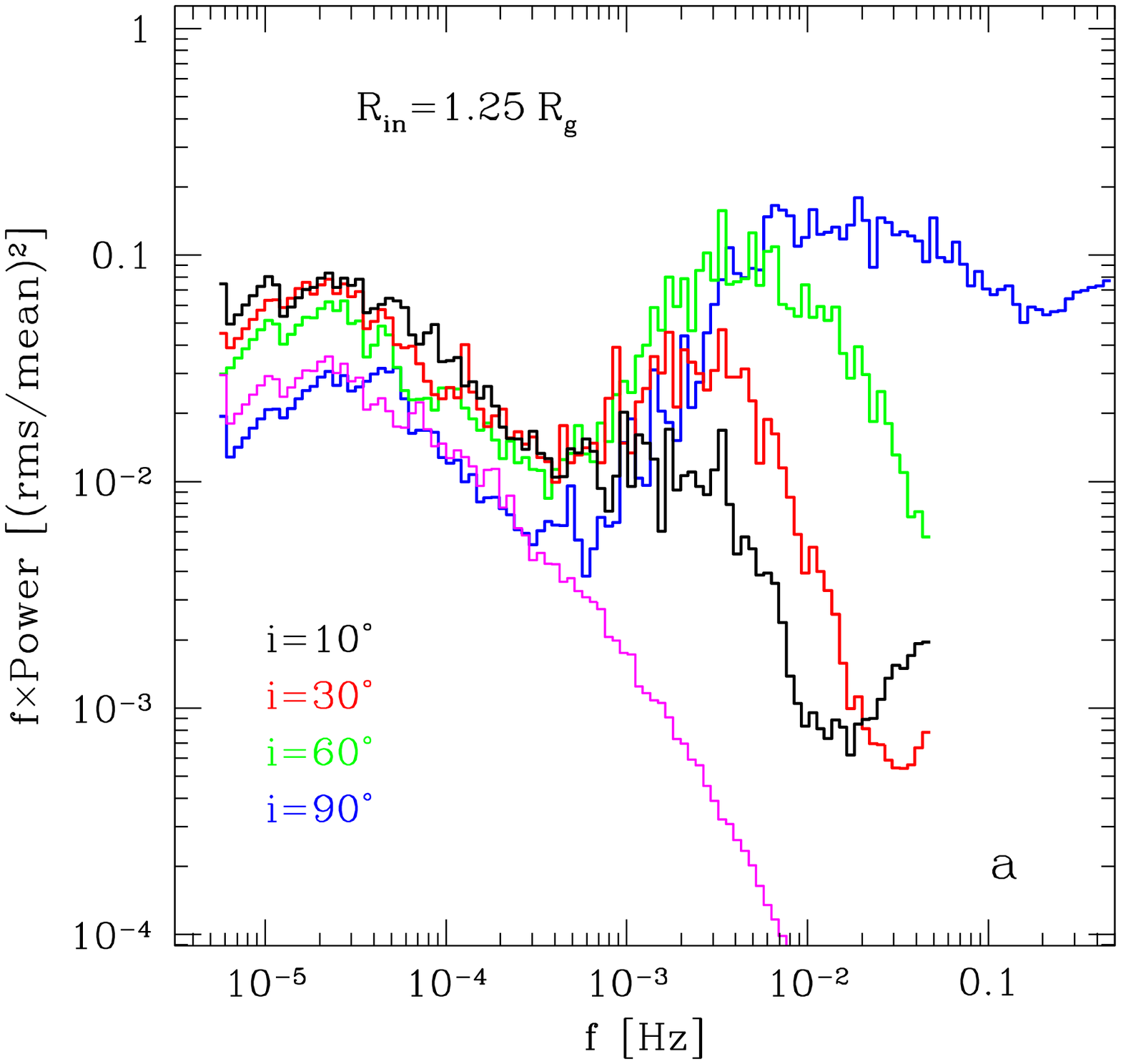}
} \hfil {
  \parbox{0.45\textwidth}{
   \epsfxsize = 0.4\textwidth
   \epsfbox[18 150 620 710]{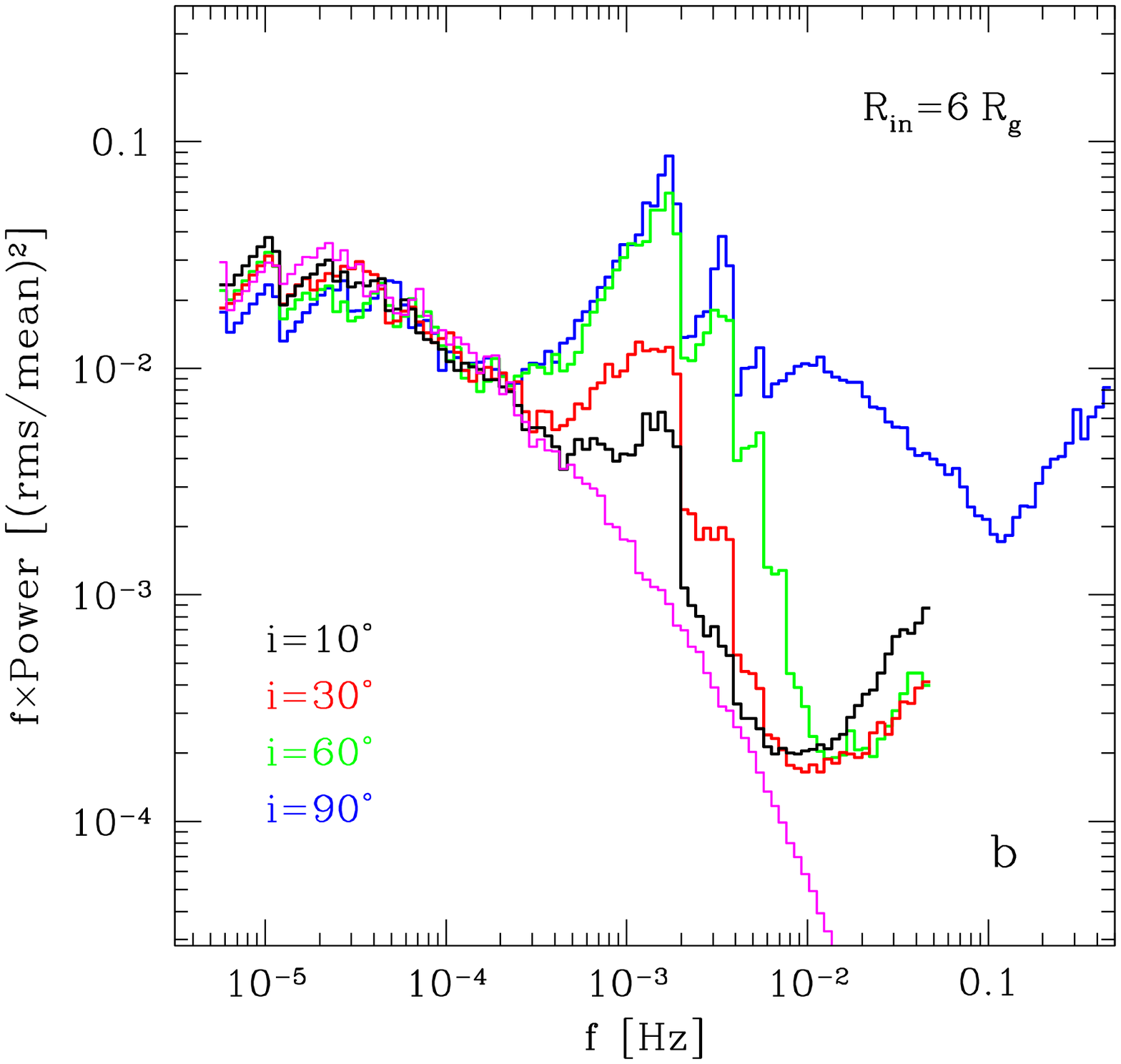}
}
}}
 \caption{
Power density spectra from the flare model, with X-ray flare emission 
modulated by relativistic effects from Keplerian motion of the flares, 
in the Kerr metric with $a=0.998$. 
The flares are concentrated toward
the inner disc radius, so that the radial emissivity follows that expected
for an accretion disc in the same metric. Inner disc radius is
$\rin=1.25\,\Rg$ (a), and $\rin=6\,\Rg$ (b). 
Results for a range of inclination angle, $i$, are plotted. 
The PDS for $i=90^\circ$ is combined from two: one from a light curve
with $\tbin=10$ s, the other with $\tbin=1$ s. The thin magenta
histograms in both panels show the PDS without the relativistic effects.
Black hole mass is $10^6\,\MSun$. For a BHB system with $10\,\MSun$
the additional signal appears above 100 Hz.
\label{fig:psd}}
\end{figure*}

The observed flux of radiation from a flare co-rotating with the disc
is periodically modulated. Main contribution to the modulations comes 
from the Doppler boost: the intensity is modulated as the observed energy 
of photons varies with their point of emission, so that the invariant
$I_{\nu}/\nu^3$ is constant. Additional effect comes from the spectral shape:
since $N_{\rm obs}(E)$ contains contributions from 
$N_{\rm em}(E/g) \propto g^{\Gamma} E^{-\Gamma}$ (where $\Gamma\sim 1.7$ is 
the photon  spectral index), this enhances the Doppler modulation.
Finally, there is also periodic modulation due to varying $\Delta {\hat t}_{\rm ph}$, 
which is influencing the times/rate of arrival of photons.

In order to demonstrate the magnitude of the modulation, we plot
in Fig.~\ref{fig:modconst} the
light curves and power spectra from an intrinsically constant signal
emitted at $1.25\,\Rg$ and $6\,\Rg$, and observed at $i=60^{\circ}$.
The modulation is very strong and non-sinusoidal. For $R_{\rm em}=1.25\,\Rg$ its 
amplitude is $\approx 2\times 10^3$. The non-sinusoidal shape of the
modulation means that the PDS contains strong harmonics -- peaks at
multiples of the fundamental frequency, $f_0 = T_{\rm orb}^{-1} = 
(2\pi/\OmegaK)^{-1}$. The case of $R_{\rm em}=6\,\Rg$ is qualitatively similar,
with the fundamental frequency smaller by factor 
$\OmegaK(6\,\Rg)/\OmegaK(1.25\,\Rg) \approx 6.5$. The underlying PDS
continuum is approximately a power law $P(f) \propto f^{-2}$, which is 
a power spectrum of a constant signal for $t>0$.

Our main results are presented in Fig.~\ref{fig:psd}. The underlying
PDS breaks from $\propto f^{-1}$ to $\propto f^{-2}$ at 
$f_{\rm br} \approx 10^{-5}$ Hz (for $10^6\,\MSun$ black hole). 
Superimposed on this continuum is a strong signal from relativistic modulation.
The signal is far from quasi-periodic, both because of the assumed range of radii
where the flares are located, and because of the strong harmonic content.
 For $\rin=6\,\Rg$ the orbital frequency
at the inner radius can actually be seen in the PDS, together with two
of the higher harmonics. The feature drops sharply at 
$f=T_{\rm orb}^{-1}(\rin) \approx 2\times 10^{-3}$ Hz, 
while its slope at low $f$ side is related
to the radial distribution of the flares (see Abramowicz et al.\ 1991).
For $\rin=1.25\,\Rg$ the relativistic signal in PDS is quite broad,
and the frequency $\OmegaK(r=\rin)$ is not evident there. The main reason for
this is again the strong harmonic content, but there are some additional
factors. Firstly, the energy dissipation rate dependence on radius, 
$Q(r)$,  goes to zero at $r=\rms\approx \rin$.
Secondly, both the correction for time units, eq.~\ref{equ:dtdt},
and the fraction of photons escaping to infinity are functions of radius,
strongly suppressing the emission from small radii.
All these combine to produce the luminosity received from a given
radius, $L(r)$, which is peaked at somewhat larger radius than $\rin$
(the peak is at $\approx 2 \,\Rg$). Also, the radial function $L(r)$
is somewhat flatter than $r^{-3}$.

There is a larger spread in the overall PDS normalization for $\rin=1.25\Rg$
than for $\rin=6\Rg$. This is most likely caused by smaller fraction of
flares (thus larger dispersion) contributing to the observed emission 
for smaller radius. The reason is again that the flares are emitted
according to the local energy dissipation rate, $Q(r)$, while in the observed
emission the contribution from smallest radii is relatively suppressed.

In the considered frequency range, $10^{-5} < f < 0.05$ Hz, the integrated 
r.m.s.\ 
is $\approx 0.09$ without the relativistic effects. The additional
integrated signal contribution to r.m.s.\ depends, obviously, on $\rin$ and $i$.
For $\rin=1.25\,\Rg$ it is 0.03--0.40 (see Fig.~\ref{fig:psd});
specifically, it is $\approx 0.06$ for $i=30^\circ$. For $\rin=6\,\Rg$
the additional variability r.m.s.\ is in the range $5\times 10^{-3}$--0.11.

For the assumed mass of $10^6\,\MSun$, the additional signal appears
between $10^{-3}$ and $10^{-2}$ Hz, i.e.\ on timescales of hundreds of
seconds. This would correspond to $f>100$ Hz for a BHB with $10\,\MSun$.

\subsection{Dependence of the strength of the relativistic signal in PDS
on radial distribution of flare time scales}
\label{sec:depend}

\begin{figure*}

 \parbox{\textwidth}{
   \parbox{0.45\textwidth}{
    \epsfxsize = 0.4\textwidth
    \epsfbox[18 150 620 710]{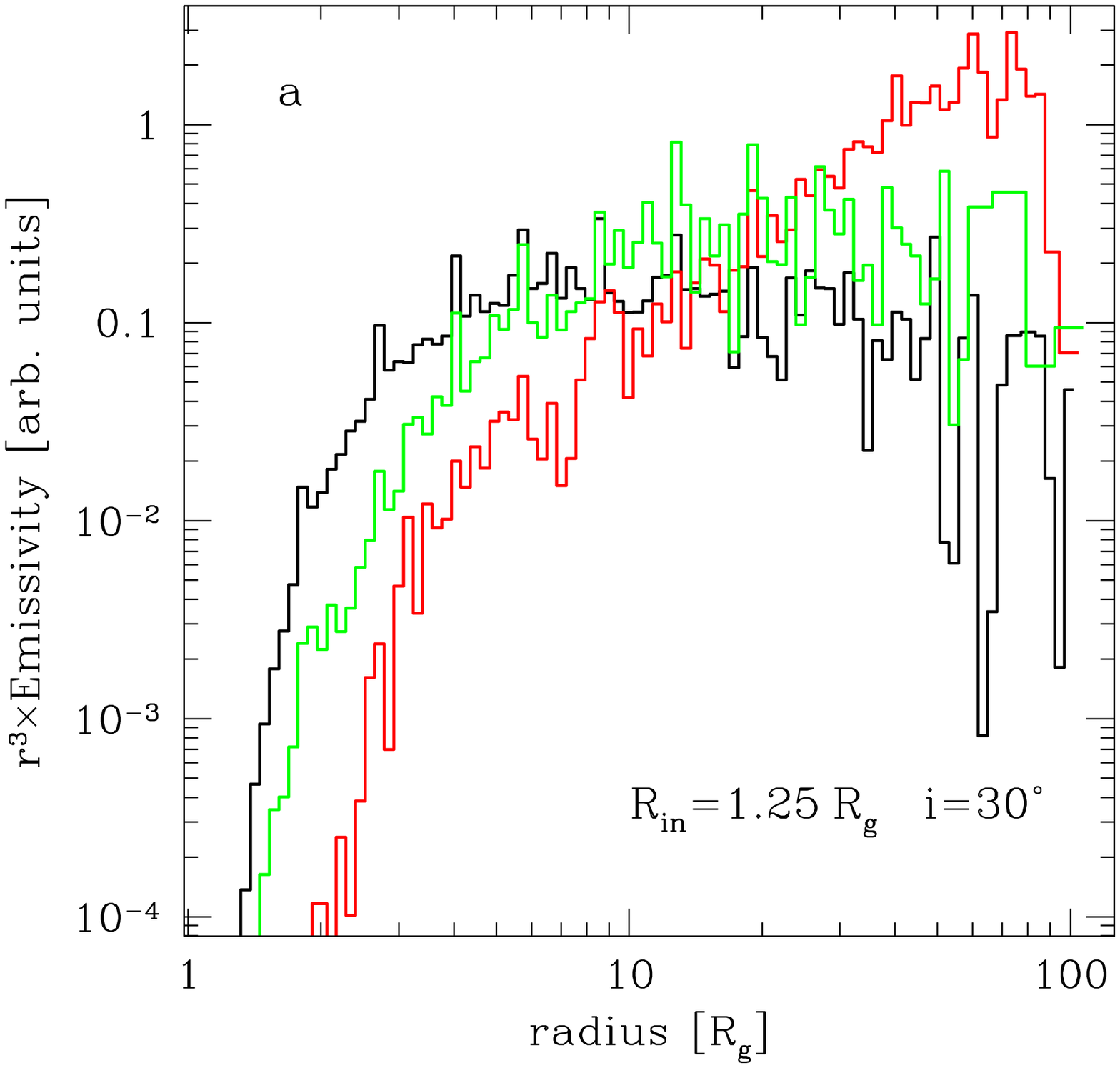}
} \hfil {
  \parbox{0.45\textwidth}{
    \epsfxsize = 0.4\textwidth
    \epsfbox[18 150 620 710]{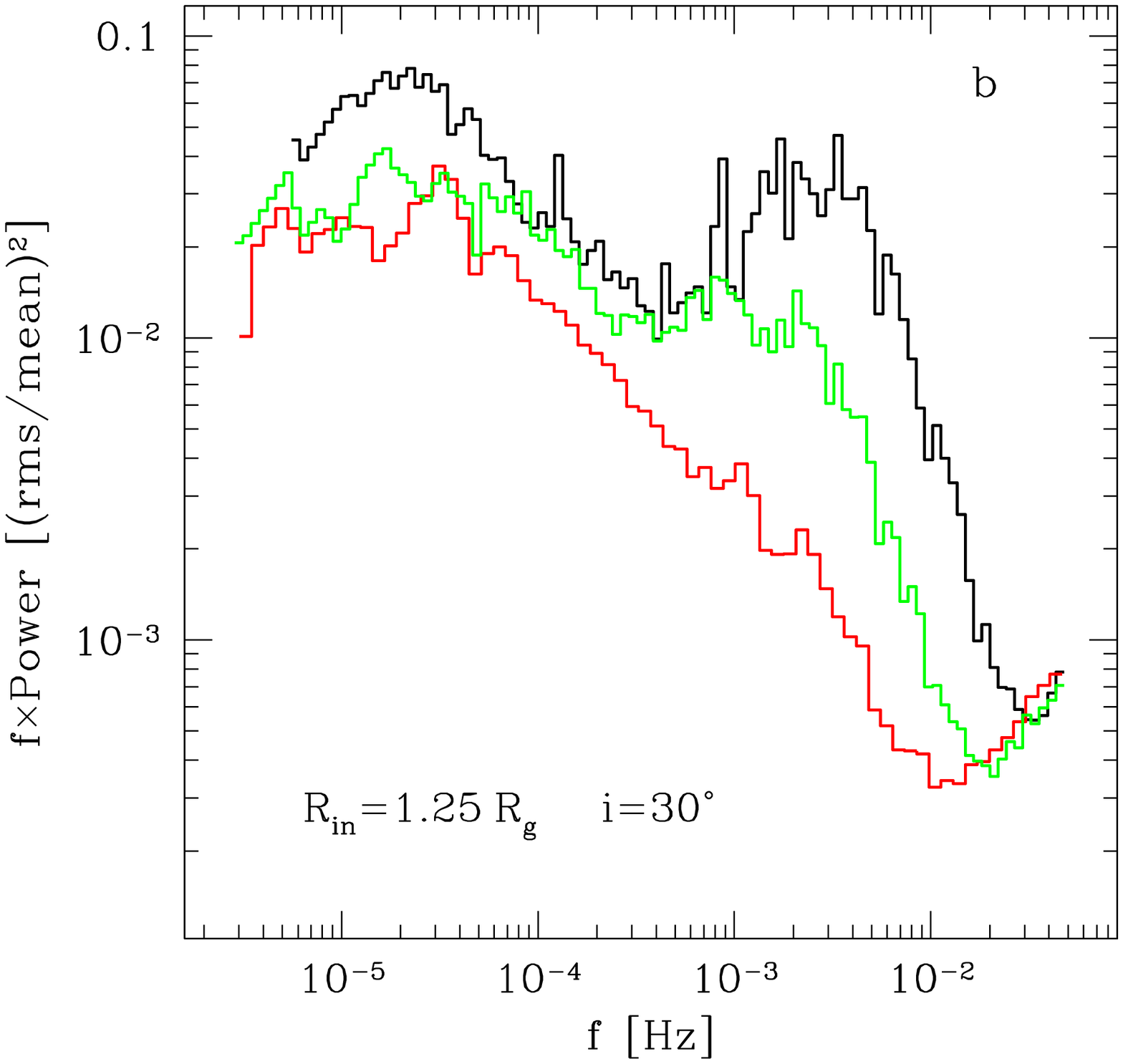}
}
}}
 \caption{
Examples of modified radial probability distribution of flares, with
resulting radial emissivities (a) and PDS (b). 
In the modified models longer flares occur preferentially at large
distance, as expected from physical arguments (see Sec~\ref{sec:depend}). 
The emissivities were multiplied by 
$r^3$ (asymptotically, $Q(r) \propto r^{-3}$). 
Black histograms show original model for $\rin=1.25\,\Rg$
and $i=30^{\circ}$ as a reference. Red histograms show a model with 
strong correlation between $r$ and $\tauf$, resulting in the emissivity peaking
at large $r$ and lack of the relativistic signal in PDS. Green histograms
demonstrate a case of relatively narrow radial distribution of short lived
flares, while the radial distribution of longer flares becomes progressively
broader. When the radial emissivity is forced to have a given steep radial
dependence, the strength of the relativistic signal in PDS is independent of
any $r$-$\tauf$ relation.
\label{fig:modrad}}
\end{figure*}

The strength of the relativistic signal in PDS may depend on assumptions
about possible correlations between radial position of a flare and its time 
scale. In our basic model there is no such relation, which means that the 
probability $\Pt(\tauf)$ is independent of radius.
The relativistic signal is reduced if longer flares  occur preferentially at
larger distance, because -- at a fixed modulation period -- the shorter the
 modulated  flare, the lower the quality factor of resulting periodic
modulation.  
We repeat, however, that the simple physically motivated prescription uniquely 
linking $r$ and
$\tauf$, $\tauf \propto \OmegaK(r)^{-1}$ leads to incorrect radial distribution
of emitted energy, which increases with radius (\.{Z}ycki 2002).  
We have therefore tested more complex $r$--$\tauf$ relations. We modified the 
radial probability distributions, multiplying the original function, $P_r(r)$, 
by a gaussian,  $\exp [-(r-r_0)^2/\sigma^2 ]$, where the center 
$r_0 \propto \tauf^{2/3}$. We have made a number of trial runs with various 
values of $\sigma$, either constant, or dependent on $r_0$. This form of modified
probability distribution, ${\tilde P}_r(r)$, indeed favours longer flares to appear
 farther away 
from the center, and the strength of this effect can be controlled by $\sigma$.
For $\sigma\rightarrow \infty$, ${\tilde P}_r(r)$ reduces to original $r Q(r)$, 
while for $\sigma\rightarrow 0$, the unique relation $r_0 \propto \tauf^{2/3}$
is recovered. Obviously, the above form of ${\tilde P}_r(r)$ is completely
arbitrary; its only purpose is to test the dependence of our results on possible
unknown relation $r$--$\tauf$.

Fig.~\ref{fig:modrad} shows examples of PDS and radial distributions of
emission for such modified probability distributions. The red histograms show
a case of strong correlation between $r$ and $\tauf$, with $\sigma=0.2 r_0$.
This results in energy emission shifted towards larger radii, and, in consequence,
no relativistic signal in PDS. The green histograms show results for
$\sigma/r_0=0.2 + 0.4 \ln(\tauf/\tmin)/\ln(\tmax/\tmin)$, meaning a narrow
radial distribution for short lived flares, which progressively becomes broader 
for longer flares. This gives the emissivity somewhat shifted to larger $r$
and weaker QPO signal compared to the basic unmodified model. However, when
the function $P_r(r)$ is now modified to be a steeper function of $r$, 
so that the radial emissivity is restored to its original dependence, $r Q(r)$, the
strength of the QPO signal is also restored (the flares are moved closer to the center).

Concluding, when the radial emissivity is forced to have a prescribed form,
the (unknown) relation between radial position and life time of a flare does not
seem to have any influence on the strength of the relativistic signal in PDS. We
do  note
that this requires that some longer flares occur close to the center, in apparent
contradiction with the basic condition that the flare time scale should be of order
of Keplerian time scale (e.g.\ Romanova et al.\ 1998).

\section{Discussion}
\label{sec:discuss}

The broad Fe \Ka lines and reflection spectra observed in X-ray data of many 
accreting black holes imply fast Keplerian rotation of the reprocessing matter, but 
they do not imply that the sources of primary radiation participate 
in this motion. Moreover, the primary spectra, being of an approximately power law 
form, are not expected to be distorted by the relativistic effects. Therefore,
we have computed timing signatures of these relativistic effects on primary
Comptonized emission. We have assumed that the basic
form of variability PDS is a result of a flare avalanche model of PF99
and that the flares correspond to compact sources co-rotating with the 
Keplerian accretion disc.
We have estimated the position and strength of the additional signal in PDS
due to periodic modulation of intensity from the Doppler and other effects. 
We now discuss briefly the applicability of our results to current data
of Seyfert galaxies and black hole binary systems.

\subsection{Seyfert galaxies} 

\begin{figure*}

 \parbox{\textwidth}{
   \parbox{0.45\textwidth}{
    \epsfxsize = 0.4\textwidth
    \epsfbox[50 200 600 680]{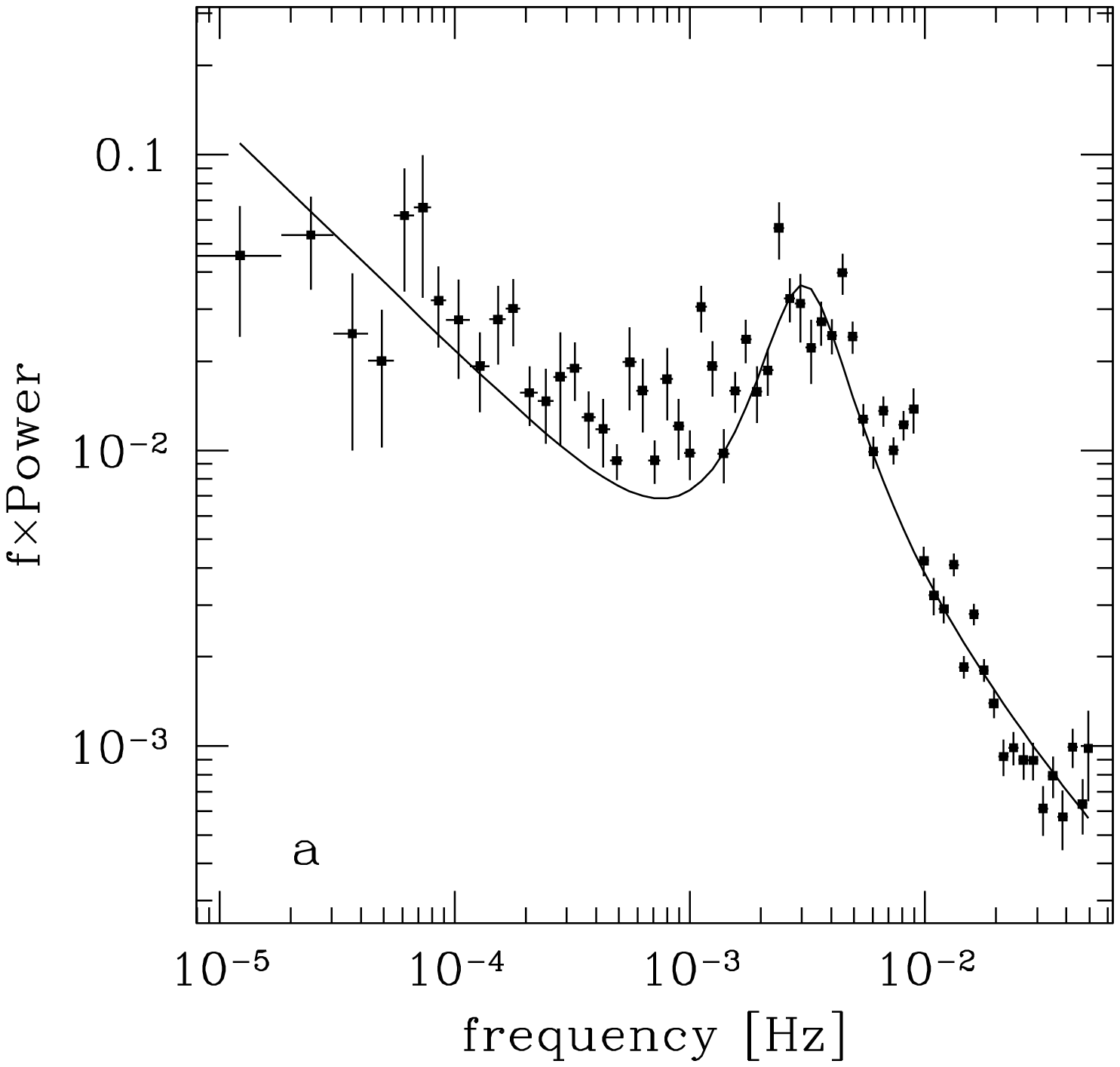}
} \hfil {
  \parbox{0.45\textwidth}{
    \epsfxsize = 0.4\textwidth
    \epsfbox[50 200 600 680]{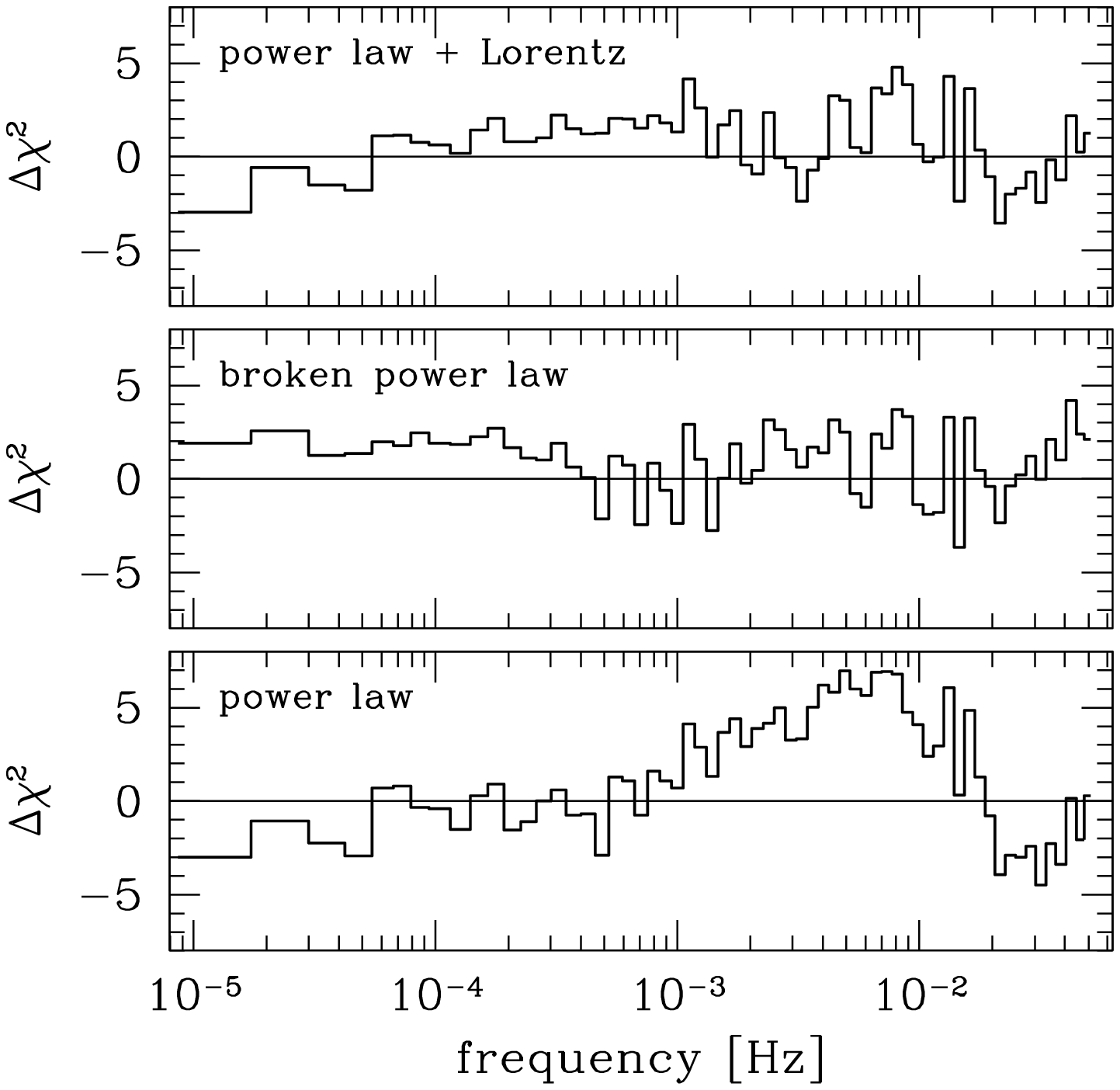}
}
}}
 \caption{
The relativistic signal could be easily seen in PDS from high quality 
{\it XMM-Newton\/} data of 
MCG--6-30-15.  The model light curve for $\rin=1.25\,\Rg$ and $i=30^{\circ}$
 (see Fig.~\ref{fig:psd}a) was used to generate a light curve of quality
corresponding to the observations of Vaughan et al.\ (2003). 
Total length of the light curve is  $\approx 4\times 10^5$ s,
the mean count rate is 20 cts/s, and the time bin is 10 sec.
Panel (a) shows the resulting PDS with Poisson level subtracted,
and with  a best fit power law plus a Lorentzian model.
The right panels show the data--model residuals. The additional signal 
in PDS is highly significant, although its detailed description
is not unique.
No such signal is seen in the actual data of that Seyfert galaxy (see fig.~4 and 
fig.~5 in Vaughan et al.\ 2003).
\label{fig:psdnoise}}
\end{figure*}

The Seyfert 1 galaxy  MCG--6-30-15 is the best example of an AGN with
a broad Fe \Ka line.
The line profile in this source apparently requires X-ray reprocessing to take
place very close to the horizon of a rapidly rotating black hole
(e.g.\ Fabian et al.\ 2002 and references therein). 
If the sources of primary
emission participate in the Keplerian emission, this would produce very clear
timing signatures. In order to see whether they would be observable in current
data, we simulate a light curve corresponding to the recent {\it XMM-Newton\/}
observations described by Vaughan et al.\ (2003). 
We use the unmodified model, i.e.\ without any $r$--$\tauf$ correlation
(Sec.~\ref{sec:depend}), since
it ensures the steep radial emissivity, as required by the Fe line profile
fits. The model light curve, $F(t)$, 
is scaled to a given mean value, $\mu$. ``Observed'' light curve is then
formed by drawing a number of counts in each time bin from Poisson distribution
of mean $F(t_{\rm n})$, where $t_{\rm n} = n\times t_{\rm bin}$, $n=1,2,\dots,N$. 
Thus simulated light curve contains the effect of Poisson noise. The time bin
is $t_{\rm bin} = 10$ s, total duration of the light curve is $\approx 10^5$ s,
while the mean value, $\mu = 20$ cts/s, all as in actual observations
(see fig.~1 in Vaughan et al.\ 2003). Fig.~\ref{fig:psdnoise} shows the resulting
PDS with the Poisson noise level subtracted, together with a best fit
power law plus Lorentzian model. A simple power law model gives a bad fit to the
data, $\chi_{\nu}^2=747/61$ d.o.f. Improvement by adding the Lorentzian is
significant, $\chi_{\nu}^2=257/58$ d.o.f., although the description of the
additional feature in PDS is not unique: a broken power law model gives a
fit of the same quality.   We have checked that similarly significant
difference between the two model fits to the PDS with relativistic effects
is maintained, even if the the mean source count rate is 4 time lower.
Thus, the signal from relativistic  modulation, if present, could be observable,
in current data. The signal is, however, not observed (see fig.~4 and fig.~5
in Vaughan et al.\ 2003). The lack of the signal is problematic for the 
model of rotating flares, suggesting alternative geometries, which do not require
Keplerian motion of the X-ray sources.

On the other hand, periodically variable redshifted  Fe \Ka line was recently reported 
in 
{\em XMM-Newton\/} data of NGC 3516 by Iwasawa, Miniutti \& Fabian (2004). 
The variability could be 
interpreted as a result of Doppler effect from a flare co-rotating with
a disc, with orbital period $\approx 25$ ksec. Since this kind of transient
modulated line emission is not expected from a model with axially symmetric
illumination, this observation provides an argument in favour of the
rotating source model. However, the primary continuum does {\em not\/} show
the expected modulation. These two observational results -- modulation of Fe \Ka
line and lack of modulation of primary continuum -- if firmly established, will
certainly require revision of current X-ray emission models.

One obvious consequence of the relativistic modulation is that highly inclined
sources should show stronger modulation than less inclined ones. This was also 
pointed out by Czerny et al.\ (2004) for Seyfert 1 vs Seyfert 2 galaxies. This
effect may not be easily observable, however,  if the geometry of accretion in
Seyfert galaxies corresponds to low state of BHB, and the accretion discs are truncated
far from the last stable orbit. Since the strength of the relativistic signal
decreases very quickly with increasing inner disc radius, these effects can only be 
observed in sources in high states where the disc extends all the way to the last stable
orbit.

\subsection{Black hole binaries/other objects}

Two BHB were reported to have very broad Fe \Ka lines in their {\em XMM-Newton\/}
spectra:  GX 339-4 (Miller et al.\ 2004) and GRO J1650-500 (Miller et al.\ 2002). 
The inner disc radius inferred from the
fits to the line profiles are 2--3$\Rg$, similar to that in MCG--6-30-15.
Thus, in principle, high-frequency PDS should contain the signal due to
relativistic modulation. However, as is well known, stellar-mass black hole 
systems are less suitable to investigating short time-scale variability than AGN:
even though their X--ray fluxes are $\sim 10^3$ times higher than from AGN,
the required time-scales are $10^5$--$10^7$ times shorter. Our simulations
indicate that it would be rather difficult to detect the additional signal
above 100 Hz. On the other hand,
high-frequency quasi-periodic oscillations (QPO) {\em are\/} observed in power
spectra of X--ray binaries. Some of those QPO are located at high frequencies,
$f>100$ Hz, i.e.\ in a region where the signal from relativistic modulations 
might appear (see, e.g., Schnittman \& Bertschinger 2004).
It is rather uncertain, however, how this mechanism might explain
the observed QPO, since it would require modulation from a narrow ring of the disc, 
and  suppression of the harmonics, in order to produce a narrow QPO feature.

\section{Conclusion}

Keplerian motion of sources of X--ray radiation in accreting black holes 
gives a very strong signature in power spectra, if the sources are located
close to the center. Such signal, if present, would be observed in recent high 
quality data, 
for example the {\it XMM-Newton\/} data from the Seyfert 1 galaxy MCG--6-30-15.
Lack of the signal in those data posses problems for models assuming that the X--ray
sources participate in the Keplerian disc motion, as for example expected for 
magnetic flares.

\section*{Acknowledgments} 
 
We thank Chris Done and Kris Beckwith for discussions on relativistic effects.
This work  was partly supported by grants no.\  2P03D01225
and PBZ-KBN-054/P03/2001
from the Polish State Committee for Scientific Research (KBN).

{}


\end{document}